\documentstyle[11pt,amssymb]{article}

\def\dalemb#1#2{{\vbox{\hrule height .#2pt
        \hbox{\vrule width.#2pt height#1pt \kern#1pt
                \vrule width.#2pt}
        \hrule height.#2pt}}}

\def\0{{\sst{(0)}}}
\def\1{{\sst{(1)}}}
\def\2{{\sst{(2)}}}
\def\3{{\sst{(3)}}}
\def\4{{\sst{(4)}}}
\def\5{{\sst{(5)}}}
\def\6{{\sst{(6)}}}
\def\7{{\sst{(7)}}}
\def\8{{\sst{(8)}}}

\def\ep{\epsilon}

\def\nn{\nonumber} \def\bd{\begin{document}} \def\ed{\end{document}}
\def\ds{\documentstyle} \let\fr=\frac \let\bl=\bigl \let\br=\bigr
\let\Br=\Bigr \let\Bl=\Bigl 
\let\bm=\bibitem
\let\na=\nabla
\let\pa=\partial \let\ov=\overline 
\newcommand{\be}{\begin{equation}} 
\newcommand{\ee}{\end{equation}} 
\def\ba{\begin{array}}
\def\ea{\end{array}}
\def\ft#1#2{{\textstyle{{\scriptstyle #1}\over {\scriptstyle #2}}}}
\def\fft#1#2{{#1 \over #2}}
\def\del{\partial}
\def\sst#1{{\scriptscriptstyle #1}}
\def\oneone{\rlap 1\mkern4mu{\rm l}}
\def\ie{{\it i.e.\ }}
\def\via{{\it via}}
\def\semi{{\ltimes}}
\def\str{{\rm str}}
\def\jm{{\rm j}}
\def\im{{\rm i}}
\def\mapright#1{\smash{\mathop{-\!\!\!-\!\!\!-\!\!\!-\!\!\!-\!\!\!
             \longrightarrow}\limits^{#1}}}
\def\maprightt#1#2{\smash{\mathop{-\!\!\!-\!\!\!-\!\!\!-\!\!\!-\!\!\!
             \longrightarrow}\limits^{#1}_{#2}}}

\newcommand{\ho}[1]{$\, ^{#1}$}
\newcommand{\hoch}[1]{$\, ^{#1}$}
\newcommand{\bea}{\begin{eqnarray}} 
\newcommand{\eea}{\end{eqnarray}} 
\newcommand{\ra}{\rightarrow}
\newcommand{\lra}{\longrightarrow}
\newcommand{\Lra}{\Leftrightarrow}
\newcommand{\ap}{\alpha^\prime}
\newcommand{\bp}{\tilde \beta^\prime}
\newcommand{\tr}{{\rm tr} }
\newcommand{\Tr}{{\rm Tr} } 
\newcommand{\NP}{Nucl. Phys. }
\newcommand{\tamphys}{\it Center for Theoretical Physics\\
Texas A\&M University, College Station, Texas 77843}
\newcommand{\ens}{\it Laboratoire de Physique Th\'eorique de l'\'Ecole
Normale Sup\'erieure\hoch{2,3}\\
24 Rue Lhomond - 75231 Paris CEDEX 05}
\newcommand{\upenn}{\it Department of Physics and Astronomy\\
University of Pennsylvania, Philadelphia, Pennsylvania 19104}

\newcommand{\auth}{H. L\"u\hoch{\dagger1} and
and C.N. Pope\hoch{\ddagger2}}

\begin{document}
\begin{flushright}
\hfill{CTP TAMU-26/99}\\
\hfill{UPR-850-T}\\
\hfill{SISSA-Ref.\ 68/99/EP}\\
\hfill{hep-th/9906168}\\
\hfill{June 1999}\\
\end{flushright}


\begin{center}
{ \large {\bf Exact Embedding of $N=1$, $D=7$ Gauged Supergravity in
$D=11$}}

\vspace{15pt}
\auth

\vspace{15pt}

{\hoch{\dagger}\upenn}

\vspace{15pt}
{\hoch{\ddagger}\tamphys\\
and\\
SISSA, Via Beirut No. 2-4, 34013 Trieste, Italy}

\vspace{40pt}

\underline{ABSTRACT}
\end{center}

   We obtain the explicit and complete bosonic non-linear Kaluza-Klein
ansatz for the consistent $S^4$ reduction of $D=11$ supergravity to
$N=1$, $D=7$ gauged supergravity. This provides a geometrical
interpretation of the lower dimensional solutions from the
eleven-dimensional point of view.

{\vfill\leftline{}\vfill
\footnoterule
{\footnotesize \hoch{1} Research supported in part by DOE grant 
DE-FG02-95ER40893 \vskip -12pt} \vskip 14pt
{\footnotesize  \hoch{2} Research supported in part by DOE 
grant DE-FG03-95ER40917.\vskip  -12pt}}

\pagebreak
\setcounter{page}{1}

     Kaluza-Klein dimensional reduction has enjoyed a resurgence of
attention in recent years, owing to the important role that it plays
in the derivation and discussion of duality symmetries in string
theories.  For example, the non-perturbative U-duality symmetries of
the compactified string \cite{ht,wit1} are understood at the level of
the effective low-energy field theory by performing Kaluza-Klein
reductions on toroidal internal spaces of various dimensions.  One
makes contact with the conjectured non-perturbative BPS configurations
of the string by showing that there exist corresponding soliton
solutions of the lower-dimensional theories, which preserve some
fraction of the supersymmetry.  A crucial aspect of the reduction
procedure is that it is {\it consistent}, which ensures that solutions
of the lower-dimensional equations of motion will also be solutions of
the original more fundamental higher-dimensional ones.

    Recently, a new duality was conjectured, relating supergravities
in anti-de Sitter (AdS) backgrounds to superconformal field theories
on the boundary \cite{mald,gkp,wit2}.  The relevant supergravities are
{\it gauged} theories, which are believed to arise from the
Kaluza-Klein reduction of string theory or M-theory on certain
spherical internal spaces.  It again becomes crucial to know whether
the reduction is a consistent one, since this is important for
establishing the details of the link between the supergravity theory
in the bulk, and the conformal field theory on the boundary.

    In the case of toroidal internal spaces, the consistency of the
reduction procedure is guaranteed by simple group-theoretic arguments.
The consistency of the Kaluza-Klein sphere reduction of supergravities
on AdS$\times$Sphere is more subtle to address, since there appears to
be no simple group-theoretic argument that guarantees it.  Indeed,
there are simple arguments which demonstrate that the reduction and
truncation of a generic theory on a sphere will definitely be {\it
inconsistent}.  In fact it is only because of very remarkable
``conspiracies'' between the various terms in the higher-dimensional
supergravity theories that the consistent truncation is possible at
all.

    The original discussions of sphere reductions in supergravity
considered only the linearised fluctuations around a fixed
AdS$\times$Sphere background \cite{duffpope,bewn}, for which no
possibility of inconsistency arises.  Indications of the remarkable
underlying structures that lead to the consistency at the full
non-linear level were seen in studies that examined leading-order
non-linear contributions \cite{dnpw}.  The exact consistent ans\"atze
for $N=2$ and $N=3$ truncations of the $N=8$ supersymmetric $S^7$
reduction of $D=11$ supergravity were obtained in \cite{pope1,pope2}.
A complete demonstration of the consistency was presented in
\cite{deWitnicolai}, although the construction was highly implicit,
and did not lend itself to the explicit re-interpretation of
lower-dimensional solutions in terms of eleven-dimensional ones.
Results for the full non-linear ansatz just for the metric in certain sphere
reductions were also obtained in \cite{wn1,wn2,wn3,nilsson}.

     Recently, the first examples of complete and explicit non-trivial
reductions were presented, for $S^7$ and $S^4$ compactifications of
M-theory and an $S^5$ compactification of the type IIB string
\cite{ten}.\footnote{We define ``non-trivial'' reductions to mean ones
that involve scalar fields that parameterise inhomogeneous
deformations of the compactifying sphere, implying that the
consistency of the reduction will not be explicable purely by a simple
group-invariance argument.  Furthermore, we emphasise the completeness
of the reduction ansatz because it is only when one has the full set
of non-linear expressions, including in particular the ans\"atze for
the higher-dimensional field strengths as well as the metric, that one
is able to obtain a consistent reduction.  The metric sector by itself
gives rise to inconsistencies that are resolved only by virtue of
``miraculous'' conspiracies between the higher-dimensional metric and
antisymmetric tensor fields.}  In each case, it was possible to obtain
explicit results by making further truncations in which only fields
associated with the Cartan subalgebras of the non-abelian Yang-Mills
gauge groups were retained.  Thus in the $S^7$, $S^4$ and $S^5$
reductions only the gauge fields of $U(1)^4$, $U(1)^2$ and $U(1)^3$
respectively were retained.  These reductions were sufficient,
however, for allowing the re-interpretation of certain non-trivial BPS
solutions of the lower-dimensional supergravities as solutions in
$D=11$ or $D=10$.  In particular, it was shown that charged AdS
black-holes in the lower dimensions could be interpreted as the
near-horizon limits of corresponding rotating M-branes or D3-brane in
$D=11$ and $D=10$ \cite{Cveticgubser1,ten}.

     The construction in \cite{ten} showed that the reduction must be
performed at the level of the higher-dimensional equations of motion,
rather than the Lagrangian (even when it exists).  Furthermore, it
also demonstrated how the consistency of the reduction depends
crucially on ``conspiracies'' involving an interplay between the
various fields in the higher-dimensional theory, and emphasised the
fact that the Kaluza-Klein reduction of a ``generic'' theory on a
sphere would, by contrast, be inconsistent.

    In this letter, we shall consider the Kaluza-Klein reduction of
eleven-dimensional supergravity on $S^4$.  A complete Kaluza-Klein
reduction ansatz for the $S^4$ compactification to $N=2$ gauged
supergravity in $D=7$ was recently presented \cite{nvv}.  Although
much simpler than the corresponding ansatz for the $S^7$ reduction in
\cite{deWitnicolai}, owing to the smaller dimension of the sphere, it
is still quite a complicated construction.  Here, we shall consider a
somewhat simpler situation, where we still reduce $D=11$ supergravity
on $S^4$, but where we truncate to the bosonic fields of $N=1$ gauged
supergravity.  These comprise the metric, a dilatonic scalar field, a
3-form potential and the gauge fields of $SU(2)$ Yang-Mills.  This is
still a non-trivial reduction, in that the scalar field parameterises
inhomogeneous deformations of the 4-sphere, and the gauge fields of a
non-abelian Yang-Mills group are retained.  In addition, the emergence
of the topological mass term for the 3-form potential can be seen.
Nonetheless, the construction is completely explicit, and any solution
of the seven-dimensional equations of motion can be straightforwardly
re-interpreted back in $D=11$.  Note that the embedding that we shall
discuss in this letter is qualitatively different from the one
presented in \cite{nvv}, in that we give our ansatz on the 
3-form potential and metric of the eleven-dimensional theory in its
original second-order formulation, rather than in the new first-order 
formalism presented in \cite{nvv}.  

   Our starting point is the bosonic sector of eleven-dimensional
supergravity, which is described by the Lagrangian \cite{cjs}
\be
{\cal L}_{11} = \hat R\, {\hat *\oneone} -\ft12 {\hat *\hat F_\4}
\wedge \hat F_\4 - \ft16 \hat F_\4\wedge \hat F_\4 \wedge 
\hat A_\3\ ,\label{d11lag}
\ee
where $\hat F_\4 = d\hat A_\3$, and we use hats to denote
eleven-dimensional fields and the eleven-dimensional Hodge dual $\hat
*$.  The equations of motion following from this Lagrangian are
\bea
&&\hat R_{MN} = \ft1{12} (\hat F^2_{MN} - \ft1{12} \hat F_\4^2 \, \hat
g_{MN})\ ,\nn\\
&& d{\hat * \hat F_\4} = \ft12 \hat F_\4\wedge \hat F_\4\ .\label{d11eom}
\eea

    We obtained the ansatz for the reduction to the bosonic sector of
$N=1$ gauged supergravity in $D=7$ by first taking the results for the
$U(1)^2$ abelian truncation given in \cite{ten}, setting the two
scalars $X_1$ and $X_2$ of that theory equal, and also the two $U(1)$ 
gauge fields.  Having done this, the metric on the internal 4-sphere takes
the form
\be
ds_4^2 = X^3\, \Delta\, d\xi^2 + \ft14 X^{-1}\, \cos^2\xi \Big( 
d\theta^2 + \sin^2\theta\, d\varphi^2 + (d\psi +\cos\theta\, d\varphi 
-g\, A_\1)^2\Big)\ ,\label{4met1}
\ee
where $X=e^{-\phi/\sqrt{10}}$ parameterises the scalar field in terms
of a canonically-normalised dilaton $\phi$, and 
\be
\Delta = X^{-4}\,\sin^2\xi + X\, \cos^2\xi\ .
\ee
At $X=1$, in the absence of the $U(1)$ gauge field $A_\1$, this metric
describes a unit 4-sphere as a foliation of 3-spheres that are
parameterised by Euler angles $(\theta,\varphi,\psi)$, with
``latitude'' coordinate $\xi$.  The 3-sphere itself is a $U(1)$ bundle
over the 2-sphere with metric $d\theta^2 + \sin^2\theta\, d\varphi^2$,
with $\psi$ as the coordinate on the $U(1)$ fibres.  When the
7-dimensional scalar $X$ is excited, it describes inhomogeneous
deformations of the 4-sphere, leaving the 3-sphere foliations intact.
Excitations of the $U(1)$ gauge field $A_\1$ in seven dimensions
describe deformations of the 3-sphere's $U(1)$ bundle.

     It is now natural to consider a non-abelian generalisation of the
deformations of the 4-sphere metric, by introducing the three
left-invariant 1-forms $\sigma^i$ on $S^3$, which satisfy $d\sigma^i =
-\ft12 \ep_{ijk}\, \sigma^j\wedge \sigma^k$.  These can be written in
terms of the Euler angles as $\sigma_1+\im\, \sigma_2 = e^{-\im\psi}\,
(d\theta + \im\, \sin\theta\, d\varphi)$, $\sigma_3=d\psi +
\cos\theta\, d\varphi$.  We are naturally led to generalise
(\ref{4met1}) to
\be
ds_4^2 = X^3\, \Delta\, d\xi^2 + \ft14 X^{-1}\, \cos^2\xi \sum_{i=1}^3
(\sigma^i-g\, A_\1^i)^2\ .\label{4met2}
\ee
Note that this reduces to (\ref{4met1}) in the abelian limit where the
$i=1$ and $i=2$ components of the $SU(2)$ gauge potentials $A_\1^i$ are
set to zero.

    Having established our notation, we can now present our ans\"atze 
for the $D=11$ metric and 3-form potential:
\bea
d\hat s^2_{11} &=& 
\Delta^{1/3}\, ds_7^2 + 2 g^{-2} \, X^3\,
\Delta^{1/3}\, d\xi^2 + \ft12 g^{-2}\, \Delta^{-2/3}\, X^{-1}\,
\cos^2\xi\, \sum_i (\sigma^i-g\, A_\1^i)^2\ ,\label{metans}\\
\hat A_\3 &=& \sin\xi\, A_\3 + \ft1{2\sqrt2} g^{-3}(2\sin\xi + \sin\xi\,
\cos^2\xi\, \Delta^{-1}\, X^{-4})\, \ep_\3\nn\\
&&-\ft1{\sqrt2} g^{-2}\, \sin\xi\, F_\2^i\wedge h^i - \ft1{\sqrt2}
g^{-1}\, \sin\xi\, \omega_\3\ .\label{aans}
\eea
where $h^i\equiv \sigma^i - g\, A_\1^i$ and $\ep_\3 \equiv h^1\wedge h^2
\wedge h^3$.  The
$SU(2)$ Yang-Mills field strengths $F_\2^i$ are given by $F_\2^i =
dA_\1^i + \ft12 g\, \ep_{ijk}\, A_\1^j\wedge A_\1^k$, and we have
defined $\omega_\3\equiv A_\1^i\wedge F_\2^i -\ft16 g\, \ep_{ijk}
A_\1^i\wedge A_\1^j\wedge A_\1^k$, so that $d\omega_\3= F_\2^i\wedge
F_\2^i$.  Note that all the fields $X$, $A_\3$ and $A_\1^i$, and the
metric $ds_7^2$, appearing on the right-hand sides of (\ref{metans})
and (\ref{aans}), are taken to depend only on the coordinates of the
seven-dimensional spacetime.  

        It is useful to present the field strength $\hat F_\4=
d\hat A_\3$ and its eleven-dimensional Hodge dual:
\bea
\hat F_\4 &=&
 -\ft1{2\sqrt2} g^{-3} (X^{-8}\, \sin^2\xi -2 X^2\,
\cos^2\xi + 3X^{-3}\, \cos^2\xi - 4 X^{-3} )\, \Delta^{-2}\,
\cos^3\xi\, d\xi\wedge \ep_\3 \nn\\
&&
-\ft5{2\sqrt2} g^{-3} \Delta^{-2}\, X^{-4}\, \sin\xi\, \cos^4\xi\,
dX\wedge \ep_\3 + \sin\xi\, F_\4\nn\\
&& +\sqrt2 g^{-1}\, \cos\xi\, X^4\,{* F_4}\wedge d\xi -
\ft1{\sqrt2} g^{-2}\, \cos\xi\, F_\2^i\wedge d\xi\wedge h^i \nn\\
&& 
-\ft1{4\sqrt2}\, g^{-2}\, X^{-4}\, \Delta^{-1}\, \sin\xi\,
\cos^2\xi\,F_\2^i \wedge h^j\wedge h^k\, \ep_{ijk}\ ,\label{fans}\\
{\hat *\hat F_\4}&=&  
-\ft1{\sqrt2} g\, (X^{-8}\, \sin^2\xi -2 X^2\,
\cos^2\xi + 3X^{-3}\, \cos^2\xi - 4 X^{-3} )\, \ep_\7 \nn\\
&&
 +5\sqrt2\, g^{-1}\, \sin\xi\, \cos\xi\, X^{-1}\, {*dX}\wedge d\xi 
+ 8 g^{-4}\, \sin\xi\, \cos^3\xi\, \Delta^{-1}\, {*F_\4}\wedge d\xi 
\wedge \ep_\3\nn\\
&&
-\ft1{2\sqrt2} g^{-3}\, \cos^4\xi\, \Delta^{-1}\, X\, F_\4\wedge
\ep_\3 -\ft1{4\sqrt2} g^{-2}\, \cos^2\xi\, X^{-2}\,
{* F_\2^i}\, \wedge h^j\wedge h^k\, \ep_{ijk} \nn\\
&&
-\ft1{\sqrt2} g^{-2}\, \sin\xi\, \cos\xi\, X^{-2}\, {* F_\2^i}\wedge
d\xi\wedge h^i\ ,
\eea
where $F_\4=dA_\3$, $*$ denotes the seven-dimensional Hodge dual
calculated in the metric $ds_7^2$, and 
$\ep_\7$ is the volume-form of $ds_7^2$.  

     A remark about the truncation that we are performing is in order
here.  What is required is the truncation to the pure $N=1$
supergravity multiplet in $D=7$.  As discussed in \cite{pnt}, the
3-form $A_\3$, which is massive, would have 20 on-shell degrees of
freedom if it satisfied an ordinary second-order field equation.
However, the 3-form field in the supergravity multiplet should have
only 10 on-shell degrees of freedom.  This is achieved by requiring
that it satisfy a {\it first order} field equation; the so-called
``odd-dimensional self-duality'' equation \cite{tpv}.  In the presence
of the $SU(2)$ Yang-Mills fields, we find that this equation will be
\be
X^{4}\, {*F_\4} = -\ft1{\sqrt2} g \, A_\3 + \ft12 
\omega_\3\ .\label{selfdual}
\ee
Since the imposition of this equation is part of our truncation
procedure, we are free to impose it, if we wish, when writing down the
expressions for the eleven-dimensional fields.  This we have done in
writing $\hat F_\4$ and $\hat {\hat *F_\4}$ above.

   We find that the equation of
motion for $\hat F_\4$, given in (\ref{d11eom}), implies the following
seven-dimensional equations:
\bea
d(X^{-1}\, {*dX}) &=& \ft15 X^4\, {*F_\4}\wedge F_\4 - \ft1{10}
X^{-2}\, {*F_\2^i}\wedge F_\2^i - \ft1{5}\, g^2\, (X^{-8} +2 X^2 -
3X^{-3}) \, \ep_\7 \ ,\nn\\
d(X^4\, {*F_\4}) &=& -\ft1{\sqrt2} g\,
F_\4 + \ft12 F_\2^i\wedge F_\2^i\ ,\label{d7eom}\\
D(X^{-2}\, {*F_\2^i}) &=& F_\4\wedge F_\2^i\ ,\nn
\eea
where $D$ denotes the gauge-covariant exterior derivative, $D\omega^i
= d\omega^i + g\, \ep_{ijk}\, A_\1^j\wedge \omega^k$.   Note that the
second-order equation for $A_\3$ here is nothing but the exterior
derivative of the first-order equation (\ref{selfdual}). 

    Substituting the ans\"atze (\ref{metans}) and (\ref{fans}) into the
eleven-dimensional Einstein equation in (\ref{d11eom}), we again
obtain the equations of motion for the scalar $X$ and the Yang-Mills
fields $A_\1^i$, as in (\ref{d7eom}), together with the
seven-dimensional Einstein equation
\bea
R_{\mu\nu} &=& 5X^{-2}\,  \del_\mu X\, \del_\nu X + \ft15 V\, g_{\mu\nu} 
  + \ft1{12} X^4\, (F^2_{\4\mu\nu} - \ft3{20} F_\4^2\, g_{\mu\nu} )\nn\\
&&+\ft12 X^{-2}\, ((F_\2^i)^2_{\mu\nu} - \ft1{10}\, (F^i_\2)^2\,
g_{\mu\nu} )\ ,\label{d7einst}
\eea
where $V$ is the scalar potential, given by
\be
V= g^2\, (\ft14 X^{-8} - 2 X^{-3} - 2 X^{2})\ .
\ee


    The above seven-dimensional equations of motion can be derived
from the Lagrangian
\bea
{\cal L}_7 &=& R\, {*\oneone} -\ft12 {*d\phi}\wedge d\phi - g^2\,( \ft14
e^{\fft{8}{\sqrt{10}}\phi} - 2 e^{\fft{3}{\sqrt{10}}\phi}
- 2 e^{-\fft{2}{\sqrt{10}}\phi})\, {*\oneone}
-\ft12 
e^{-\fft4{\sqrt{10}}\phi}\, {*F_\4}\wedge F_\4\nn\\ 
&&-\ft12 e^{\fft2{\sqrt{10}}\phi}\, {*F_\2^i}\wedge F_\2^i 
 +\ft12 F_\2^i\wedge F_\2^i \wedge A_\3 - \ft1{2\sqrt2} g\, F_\4\wedge
A_\3\ ,\label{d7lag}
\eea
where we have replaced $X$ by $X=e^{-\phi/\sqrt{10}}$, together with
the self-duality condition (\ref{selfdual}), which is to be imposed
after having obtained the equations of motion from (\ref{d7lag}).
This set of equations of motion precisely describe the bosonic sector
of $N=1$ gauged supergravity in seven dimensions.  The 
Lagrangian (\ref{d7lag}) was also given in \cite{tv}.

    The fact that we have been able to extract seven-dimensional
equations of motion by substituting (\ref{metans}) and (\ref{fans})
into the eleven-dimensional equations of motion demonstrates that our
reduction ans\"atze are consistent.  In other words, the
eleven-dimensional equations of motion are satisfied by all the field
configurations given in (\ref{metans}) and (\ref{fans}), provided that
the seven-dimensional fields satisfy the equations of motion of $N=1$
gauged supergravity. It is worth emphasising that the ability to extract 
seven-dimensional equations of motion in a consistent manner depends 
crucially on the interplay between the contributions of the three terms
in the eleven-dimensional Lagrangian (\ref{d11lag}), which leads to a
precise matching, and hence factoring-out, of the dependence on the 
coordinates of the internal 4-sphere in the eleven-dimensional equations
of motion.  In particular, the specific value of the coefficient of
the $\hat F_\4\hat F_\4\hat A_\3$ term in (\ref{d11lag}) is crucial.  
Since the magnitude of this term is also uniquely determined by 
supersymmetry, the calculations that we have presented here again 
support the notion that the consistency of non-trivial sphere reductions
is closely related to supersymmetry \cite{zilch}.  

     Our approach to finding the correct ansatz, and verifying that it
gives a consistent reduction, has been to substitute it into the
eleven-dimensional equations of motion, and then to verify that these
are satisfied provided the lower-dimensional equations of motion are
satisfied.  This, by definition, is what one means by consistency.  An
alternative approach that is sometimes discussed is to substitute the
ansatz into the higher-dimensional Lagrangian, and integrate over the
coordinates of the internal space.  Of course before doing this, one
would first have to find an independent argument for why the ansatz
was a consistent one.  (As is well known, the mere fact that one
obtains a sensible-looking lower-dimensional action by substituting an
ansatz into the higher-dimensional one does not, of itself, guarantee
that the ansatz is consistent.)  Since we have constructed an explict 
consistent reduction ansatz in this letter, it is of interest to see
what would happen if one were to substitute it into the
eleven-dimensional action.

    First, let us simplify the discussion by restricting just to the
gravity plus scalar sector of the seven-dimensional theory.
Substituting the ans\"atze (\ref{metans}) and (\ref{aans}) into the
eleven-dimensional Lagrangian (\ref{d11lag}), we find 
${\cal L}_{11} = \ft12 g^{-4}\, \cos^3\xi\, (R-\ft12
(\del\phi)^2 + W)\, \sqrt{g_3}\, \sqrt{-g_7}$, where
\bea
W &=& \ft1{12} g^2\, \Delta^{-2}\, X^{-11}\, \Big( -2\sin^2\xi (1-21
\sin^2\xi) -(7-66\sin^2\xi + 51\sin^4\xi)\, X^5 \nn\\
&&\qquad\qquad\qquad\quad\quad
 + 4\cos^2\xi\, (7+3 \sin^2\xi)\, X^{10} + 24\cos^4\xi\,
X^{15} \Big)\,.
\eea
Integrating over $\xi$, one indeed obtains the correct gravity plus scalar
sector of the seven-dimensional Lagrangian (\ref{d7lag}).\footnote{By
contrast, in an analogous calculation for the gravity plus scalar
sector of the $S^4$ reduction of the massive type IIA theory, the
corresponding function $W$ becomes singular at the limit of the $\xi$
integration, and the procedure of substituting the ansatz into the
action is problematical 
even in the scalar sector \cite{clp}.}  However, once the
gauge fields are included in the reduction ansatz, the substitution of
(\ref{metans}) and (\ref{aans}) into the eleven-dimensional action
will not give rise to the seven-dimensional action (\ref{d7lag}).  All
these defects are avoided by sticking with the equations of motion.

   In this letter, we have presented an explicit and consistent fully
non-linear reduction ansatz for the compactification of
eleven-dimensional supergravity on $S^4$, with a truncation to $N=1$
gauged supergravity in $D=7$.  In particular, we have shown how the
dilaton $\phi$ parameterises inhomogeneous deformations of the
4-sphere that leave the foliating 3-spheres undistorted.  On the other
hand, the $SU(2)$ gauge fields that form the surviving subgroup of the
$SO(5)$ Yang-Mills fields of the maximal gauged theory are associated
with deformations that correspond to right-translations under the
$SU(2)$ that acts on the 3-spheres.  Using this geometrical embedding
of the seven-dimensional theory in the eleven-dimensional one, it is
now possible to re-express any solution of the seven-dimensional $N=1$
gauged theory as a solution in the low-energy limit of M-theory.

\end{document}